\documentclass[twocolumn, prl,showpacs,superscriptaddress]{revtex4}

\usepackage{mathrsfs} 
\usepackage{amssymb} 
\usepackage{graphicx}
\usepackage{subfigure}
\usepackage{dcolumn}
\usepackage{bm}

\begin{document}

\title{Electric Field Effects on the Optical Vibrations in AB-Stacked Bilayer Graphene}

\author{R. Stein, D. Hughes, and Jia-An Yan}


\email{jiaanyan@gmail.com}
\affiliation{Department of Physics, Astronomy, and Geosciences, Towson University, 8000 York Road, Towson, MD 21252 USA}

\date{\today}

\begin{abstract}
Using first-principles methods, we show that an applied perpendicular electric field $E$ breaks the inversion symmetry of AB-stacked bilayer graphene (BLG), thereby slightly mixing the two in-plane high-energy optical vibrations ($E_g$ and $E_u$ modes). The mixed amplitudes increase parabolically with respect to the field strength when $E$$<$2.0 V/nm, and then exhibit linear dependence when $E$$>$2.0 V/nm. In contrast, the mixing effect on the out-of-plane vibrations ($A_{1g}$ and $A_{2u}$ modes) is found to be much stronger, with the mixed amplitudes nearly an order of magnitude larger than those for the in-plane modes. For the two in-plane modes, we then calculate their phonon linewidths and frequency shifts as a function of the electric field as well as the Fermi level. Our results reveal delicate interplay between electrons and phonons in BLG, tunable by the applied fields and charge carrier densities.

\end{abstract}

\pacs{63.20.kd; 63.22.Rc; 73.22.Pr; 78.30.Na}

\maketitle




AB-Stacked Bilayer graphene (BLG) is a unique platform that both the charge carrier densities (i.e., the Fermi level $E_F$) and the band structure can be tuned through applied dual-gate electric fields \cite{Ohta2006, Oostinga2007,Zhang2009}. With an applied external electric field (EEF) of several V/nm, an energy gap opening of up to 250 meV has been reported in gated BLG \cite{Zhang2009}. Recently, the development of the electrolytic gate allows the charge carrier densities in BLG doping as high as $|n|$ $\approx$ 2.4$\times$ 10$^{13}$ cm$^{-2}$ \cite{Efetov2011}. The tunability on the band structure and the Fermi level reveals rather intriguing properties in this system, including the unusual high field transport \cite{Taychatanapat2010}, the renormalization of the phonon energy \cite{Marlard2008,Yan2008,Das2009}, and the Fano resonance in infrared spectra \cite{Kuzmenko2009,Tang2010}. Interesting device application such as hot electron bolometer \cite{Yan2011} has also been reported.

Understanding the response of material's physical properties to external perturbations is crucial for its practical applications. In particular, EEF affects both electronic structure and lattice dynamics, and plays a key role on the performance of electronic devices. Although much attention has been paid to the field effects on the electronic properties of BLG \cite{Ohta2006,Zhang2009,Marlard2008,Li2009,Mak2009}, very few studies \cite{Ando2007,Ando2009} have been conducted on the response of the lattice dynamics to the EEF. Using analytical methods, Ando and Koshino \cite{Ando2009} studied the effects of an uneven charge doping and external electric fields on the self energy of the in-plane modes in BLG. On the other hand, it has been proposed more than forty years ago that EEF could be applied to break the symmetry of specific silent mode in SrTiO$_3$ so that it becomes Raman active \cite{Worlock1967,Klukhuhn1970,Brillson1971}. The electric field induced Raman scattering allows study of infrared-active and silent modes in crystals of high symmetry, thus probing the electronic properties of the system. Recent development of the tip enhanced Raman spectroscopy \cite{Berweger2009} provides controlled local probe of the response under a nonuniform electric field. However, it is unclear whether such an external perturbation will alter the phonon mode eigenvector itself or not, and to what extent the phonon wave function will be changed, if any.

Raman and infrared (IR) spectroscopy are powerful tools to probe the lattice dynamics of BLG \cite{Marlard2008,Yan2009,Li2009,Mak2009,Das2009,Gava2009,Bruna2010}. A recent IR measurement of gated BLG identified large phonon linewidth ($\sim$30 cm$^{-1}$) for the mode of frequency $\omega$ = 1587 cm$^{-1}$, which has been ascribed to the in-plane $E_u$ mode \cite{Kuzmenko2009}, while another IR work \cite{Tang2010} ascribed the observed mode to the $E_g$ mode because of broken symmetry. A possible optical mode mixing induced by EEF is also proposed to account for the splitting of the G-band in the Raman measurement in BLG \cite{Yan2009,Gava2009}. Clearly, a correct explanation of the experimental observations requires a detailed knowledge of the response of these phonon modes to the EEF from first-principles. Such knowledge would also provide a necessary insight into the phonon, electron-phonon and their effects on the transport performance of device made of BLG \cite{Yao2000,Lichtenberger2011}.


\begin{figure*}[tbp]
\centering
   \includegraphics[height=12cm,angle=-90,clip]{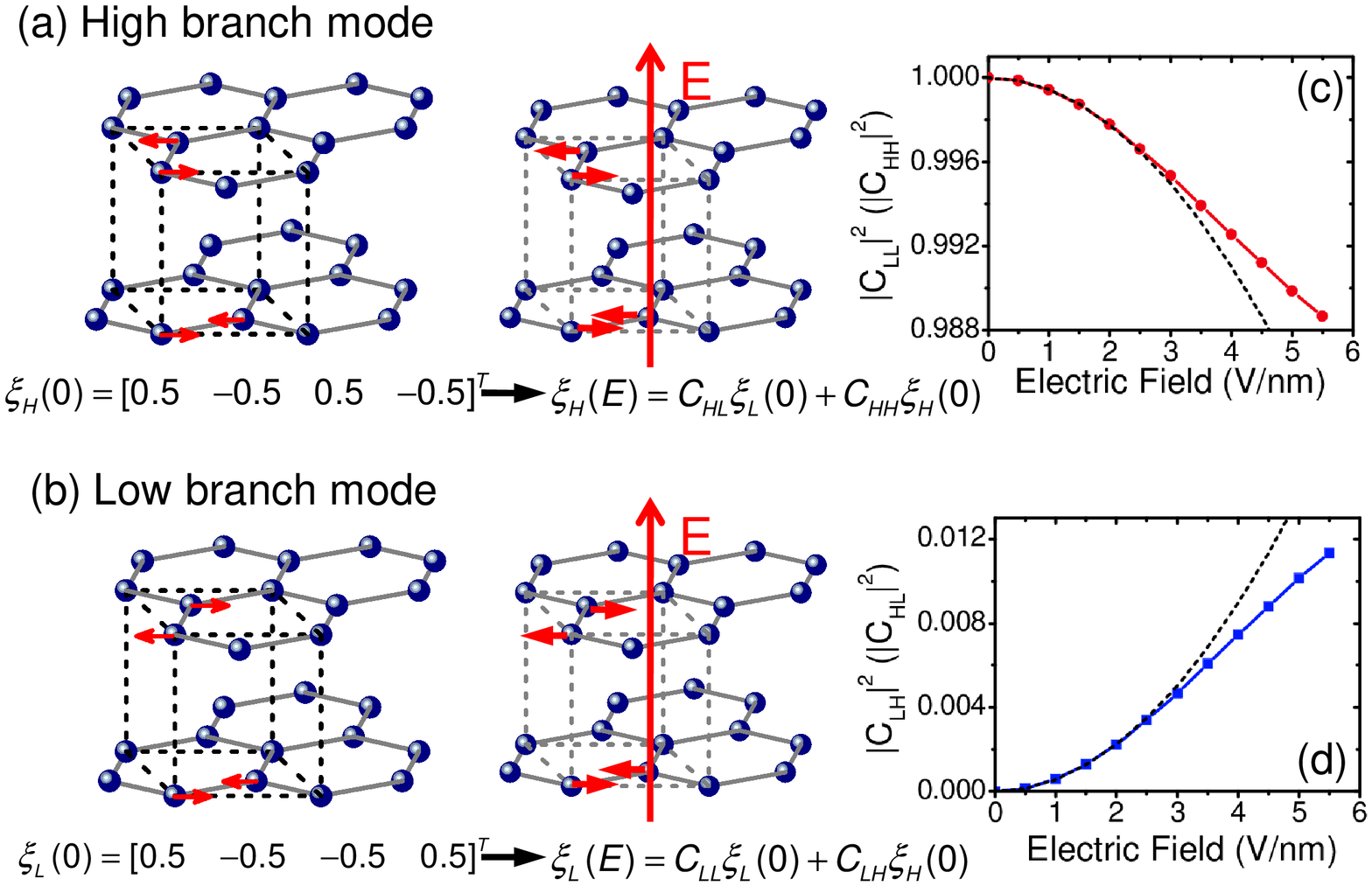}
 \caption{(Color online) Schematic plots of the in-plane (a) high-branch mode with $\omega_H$ = 1592 cm$^{-1}$
 and (b) low-branch mode with $\omega_L$=1587 cm$^{-1}$ under zero and finite electric fields, respectively. The calculated modes $\xi_L(E)$ and $\xi_H(E)$ under field can be decomposed using $\xi_L(0)$ and $\xi_H(0)$ as a complete basis set. The amplitude of (c) $|c_{LL}|^2$ ($|c_{HH}|^2$) and (d) $|c_{HL}|^2$ ($|c_{LH}|^2$) changes with respect to the field. The dashed lines are parabolic fitting of the data as $E<2$ V/nm. }\label{fig:in-plane}
\end{figure*}

\begin{figure}[tbp]
\centering
  \includegraphics[height=7.5cm,angle=-90]{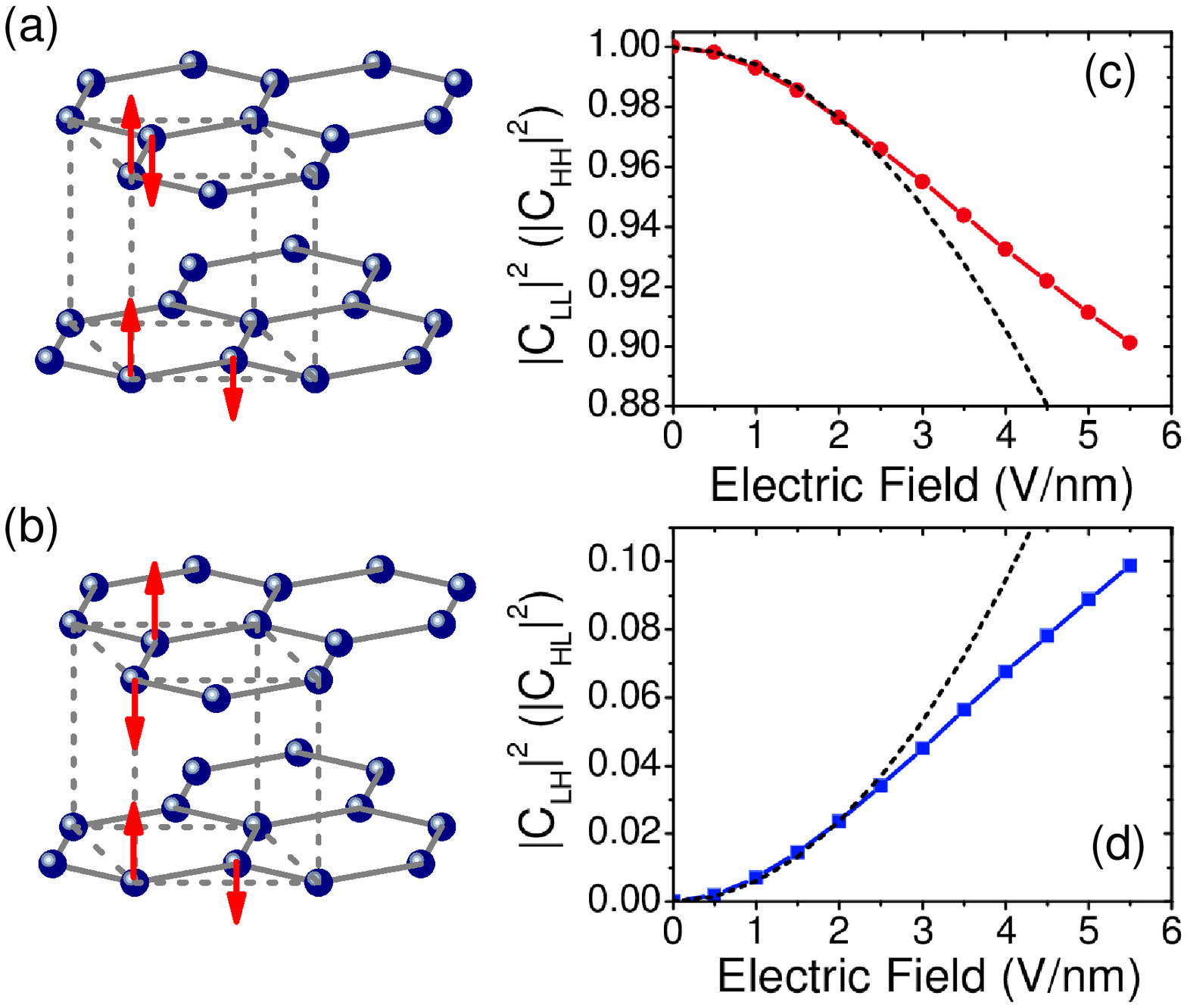}
 \caption{(Color online) Schematic plots of the out-of-plane (a) high-branch $A_{2u}$ mode with $\omega$ = 894 cm$^{-1}$
 and (b) low-branch $A_{1g}$ mode with $\omega$ = 891 cm$^{-1}$ under zero electric field. The calculated modes $\xi_L(E)$ and $\xi_H(E)$ under field can be decomposed using $\xi_L(0)$ and $\xi_H(0)$ as a complete basis set. The amplitude of (c) $|c_{LL}|^2$ ($|c_{HH}|^2$) and (d) $|c_{HL}|^2$ ($|c_{LH}|^2$) changes with respect to the field. The dashed lines are parabolic fitting of the data as $E<2$ V/nm.  }\label{fig:mode-mixing}
\end{figure}

In this work, we present a first-principles study of the electric field effects on the optical vibrations in BLG. The optical phonon modes with $q=0$ are calculated using the density-functional perturbation theory (DFPT) \cite{Baroni2001} as implemented in the Quantum ESPRESSO code \cite{pwscf} at the local density approximation (LDA) level. Norm-conserving pseudopotential \cite{Troullier1991} for carbon has been adopted to describe the core-valence interactions. The wave functions of the valence electrons are expanded in plane waves with a kinetic energy cutoff of 70 Ry. A vacuum region of 20 \AA~ has been introduced to eliminate the artificial interaction between neighboring supercells along the $z$ direction. The relaxed C-C bond length is 1.42 \AA~  and the interlayer distance is 3.32 \AA~ for the BLG. Details have been presented in our previous work \cite{me2008,me2009}. To mimic the electric field, a sawtooth potential is applied to the bilayer graphene sheets perpendicularly. More realistic configurations, such as substrate-induced electric field \cite{Ziegler2011}, should not change the physics significantly because of similar band gap opening. Dipole corrections have been considered throughout all calculations below.

Without EEF, BLG possesses a point group symmetry of $D_{3d}$. The two in-plane optical modes at $q$ = 0 are low-branch $E_g$ ($\omega$  = 1587 cm$^{-1}$) and high-branch $E_u$ ($\omega$ = 1592 cm$^{-1}$) modes, as schematically shown in Figs.~1(b) and 1(a), respectively. Depicted in Figs.~2(b) and 2(a) are the two out-of-plane optical modes $A_{1g}$ ($\omega$ = 891 cm$^{-1}$) and $A_{2u}$ ($\omega$ = 894 cm$^{-1}$) modes, respectively. We have calculated the phonon frequencies under various finite fields using DFPT. The variations of $\omega$ are within 2 cm$^{-1}$ for electric field up to 5 V/nm, indicating negligible field effects on the phonon frequency shift within the adiabatic approximation, in agreement with previous work \cite{Pisana2007}.

We then consider the phonon mode mixing induced by EEF. The applied EEF perpendicular to BLG will break the inversion symmetry of BLG, and the point group is reduced to $C_{3v}$. The two in-plane modes now belong to the E representation of $C_{3v}$, and both of them become Raman and IR active. Since E is a two-dimensional representation, the two orthonormal eigenvectors of these modes would form a complete set for the mixed in-plane modes.  Because the frequencies of the $E_g$ and $E_u$ modes are close to each other, a finite electric field will be very likely to alter the phonon wavefunctions. Hereafter the (mixed) $E_g$ and $E_u$ modes are denoted as low-branch (L) and high-branch (H) modes, respectively, as shown in Fig.~1.

Quantitatively, the mixed amplitudes can be obtained by decomposition: $\xi_H(E)=c_{HL}\xi_L(0)+c_{HH}\xi_H(0)$, and $\xi_L(E)=c_{LL} \xi_L(0)+c_{LH}\xi_H(0)$, with $c_{\alpha\beta}=\xi_\alpha^\dagger(E) \xi_\beta(0)$ ($\alpha,\beta = L, H$). The eigenvectors \cite{note} $\xi_L(0)=[0.5, -0.5, 0.5, -0.5]^T$, and $\xi_H(0)=[0.5, -0.5, -0.5, 0.5]^T$ correspond to the $E_g$ and $E_u$ modes under zero electric field, respectively. The expansion coefficients satisfy $|c_{HL}|^2+|c_{LL}|^2$ = $|c_{LH}|^2+|c_{HH}|^2$ =1, and $|c_{HL}|^2$ =$|c_{LH}|^2$. For a given electric field $E$, the amplitude of $|c_{LH}|^2$ ($|c_{HL}|^2$) shows the probability another mode contributes to the phonon wave function and is an indication of the mixing effect.

Figures 1(c) and 1(d) show the calculated $|c_{LL}|^2$ and $|c_{LH}|^2$ as a function of $E$, respectively. When $E <$ 2 V/nm, the mixed component $|c_{LH}|^2$ increases parabolically and a polynomial fitting yields $|c_{LH}(E)|^2$ = 5.6$\times$10$^{-4}E^2$ with $E$ in V/nm. When $E >$ 2 V/nm, $|c_{LH}|^2$ increases linearly with respect to $E$. The slope is found to be 2.7$\times$10$^{-3}$. Overall, the mixed component is within 1.2\% with $E$ up to 5 V/nm, implying a small mixing effects induced by EEF on the in-plane optical modes. In Ref.~[\onlinecite{Ando2009}], Ando and Koshino analyzed the phonon Green's functions of in-plane modes in BLG and reported that these modes are strongly mixed by the asymmetrical potential difference arising from bottom gate doping and external electric field \cite{Ando2009}. To clarify the effects of the charge doping and the external electric fields, we performed further calculations to investigate the phonon eigenvectors for a charged BLG under an external electric field. Our results show that the phonon mode eigenvectors are still only slightly mixed (smaller than 1\%) by the external electric field along with the charge doping.

Similar analyses can be applied to the out-of-plane modes, the low-branch $A_{1g}$ and the high-branch $A_{2u}$ modes. In Figs.~2(c) and 2(d), the amplitude of the mixed components $|c_{LL}|^2$ and $|c_{LH}|^2$ are presented as a function of $E$, respectively. When $E <$ 2 V/nm, the mixed component $|c_{LH}|^2$ increases parabolically as $|c_{LH}(E)|^2$ = 5.9$\times$10$^{-3}E^2$ with $E$ in V/nm. When $E >$ 2 V/nm, $|c_{LH}|^2$ increases linearly with a slope of 2.2$\times$10$^{-2}$. From Figs.~2(c) and 2(d), EEF has dramatic effects on the mixing of the two out-of-plane modes; the mixing components reach 10\% as $E$ = 5.5 V/nm. This is nearly an order of magnitude larger than that for the in-plane modes. It was previously assumed that the electric field has negligible effects on the phonon eigenmodes \cite{Worlock1967}. In contrast, as demonstrated here, the out-of-plane phonon modes in layered systems could be significantly altered by EEF.

We then turn to the electron-phonon coupling (EPC) of the two in-plane modes. EPC plays a key role in understanding many phenomena \cite{Marlard2008,Ando2007,Yan2008,Kuzmenko2009,Tang2010}, especially the Raman frequency shift and broadening. The phonon self-energy $\Pi_{\mathbf{q}\nu} (\omega)$ of a phonon with wave vector $\mathbf{q}$, branch index $\nu$, and frequency $\omega_{\mathbf{q}\nu}$ provides the renormalization and the damping of that phonon due to the interaction with other elementary excitations. Following the Migdal approximation, the self-energy induced by the EPC in BLG reads \cite{Ando2007, me2012}:
\begin{widetext}
\begin{equation}
\Pi_{\mathbf{q}\nu}(\omega)
=
2 \sum_{mn}\int\frac{d\mathbf{k}}{\Omega_{\mathrm{BZ}}}|g_{mn}^{\nu}(\mathbf{k},\mathbf{q})|^2 \frac{[f(\epsilon_{n\mathbf{k+q}})-f(\epsilon_{m\mathbf{k}})][\epsilon_{n\mathbf{k+q}}-\epsilon_{m\mathbf{k}}]}{(\epsilon_{n\mathbf{k+q}}-\epsilon_{m\mathbf{k}})^2-(\hbar\omega+i\eta)^2},
\end{equation}
\end{widetext}
where $\epsilon_{m\mathbf{k}}$ is the energy of an electronic state $|m\mathbf{k}\rangle$ with crystal momentum $\mathbf{k}$ and band index $m$, $f(\epsilon_{m\mathbf{k}})$ the corresponding Fermi occupation, and $\eta$ is a positive infinitesimal. A factor of 2 accounts for electron spin. For a given mode $\omega=\omega_0$, the phonon linewidth is $\gamma$ = $-2\mathrm{Im}(\Pi_{\mathbf{q}\nu} (\omega_0))$ and the phonon frequency shift is $\Delta \omega$ = $\frac{1}{\hbar}[\mathrm{Re}(\Pi_{\mathbf{q}\nu} (\omega_0)|_{E_F}-\Pi_{\mathbf{q}\nu}(\omega_0)|_{E_F=0}]$.

The EPC matrix elements in Eq.~(1) are calculated using the frozen-phonon approach developed in our previous work \cite{me2009}. We have considered the cases with and without the mixing effects. Only a negligible difference has been noticed since the mixed amplitude is within 1.2\% for the range of the electric fields considered. Below we only show the results with mixing effects included. Calculations of the phonon self-energy have been carried out on a dense 101$\times$101 $k$-grid within a minizone (0.2$\times$0.2) enclosing the BZ corner $K (K')$ in the reciprocal space. This is equivalent to 500$\times$500 $k$-grid sampling in the whole Brillouin zone. By changing the Fermi level $E_F$ in Eq.~(1), the dependence of $\gamma$ and $\Delta \omega$ on different doping levels can be investigated, assuming the EPC matrix elements are unchanged. This approximation is justified by the small dependence of the EPC matrix elements on doping for the $\Gamma$ phonon modes in graphene \cite{Attaccalite2010}. For all the linewidths calculated below, a parameter of $\eta$ = 5 meV has been used.

\begin{figure}[tbp]
\centering
  \includegraphics[width=8.5cm]{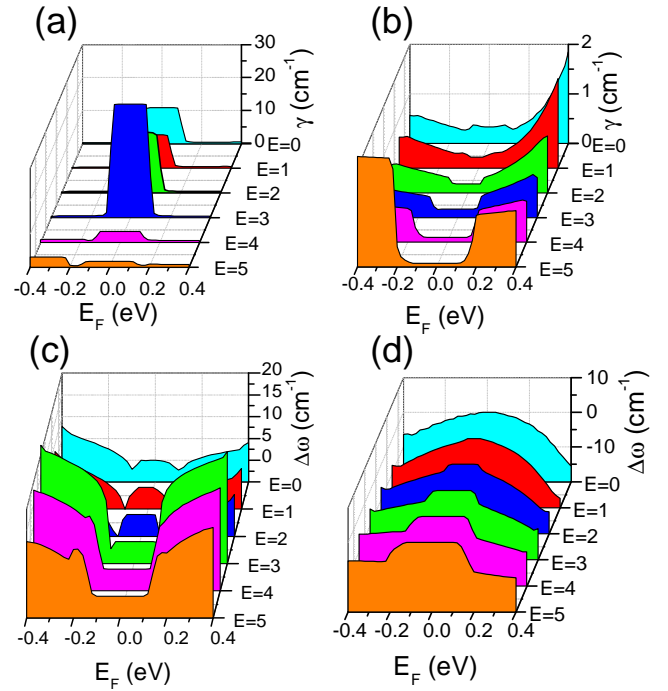}
 \caption{(Color online) Calculated phonon linewidth $\gamma$ of the (a) low-branch and (b) high-branch in-plane modes as a function of the Fermi level $E_F$ as well as the electric field $E$ (in V/nm). The corresponding frequency shifts $\Delta \omega$ are shown in (c) and (d), respectively. The neutrality point has been shifted to zero. }\label{fig:gamma}
\end{figure}

\begin{figure}[tbp]
\centering
  \includegraphics[width=8.5cm]{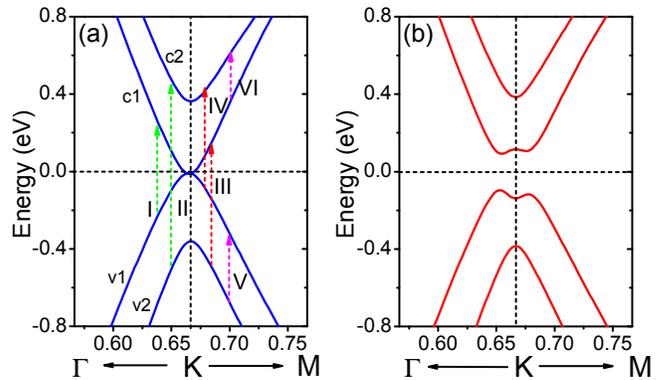}
 \caption{(Color online) Band structure of BLG under (a) zero electric field and (b) electric field of $E$ = 3 V/nm. As indicated in (a), the low-branch $E_g$ mode couples with transitions I (v1-c1), II (v2-c2), while the high-branch $E_u$ mode couples with transitions of III, IV, V, and VI. The neutrality point has been shifted to zero. }\label{fig:band}
\end{figure}

Figures~\ref{fig:gamma}(a) and~\ref{fig:gamma}(b) show the calculated linewidth $\gamma$ for the low-branch ($E_g$) and high-branch ($E_u$) modes as a function of the Fermi level $E_F$ and the electric field strength $E$, respectively. When the system is neutral ($E_F$ =0), the linewidth $\gamma$ of the $E_g$ mode increases from 9.0 cm$^{-1}$ to 10 cm$^{-1}$ as $E$ increases from zero to 1 V/nm. It increases to the maximum of 30 cm$^{-1}$ when $E$ = 3 V/nm. After $E$ $>$  3 V/nm, $\gamma$ decreases to zero rapidly. As shown in Fig.~\ref{fig:band}(b), the largest $\gamma$ at $E$ =3 V/nm is due to the fact that the electronic transition amplitude is the highest when the field-induced band gap $\sim$ 0.2 eV, close to the phonon energy. In contrast, when $E_F$ =0, the high branch $E_u$ mode exhibits nearly negligible linewidth for all the electric fields. This result clearly shows that only the low-branch $E_g$ mode can possibly exhibit a large linewidth as observed in previous IR measurement \cite{Kuzmenko2009}, and thus resolved the discrepancy on the mode assignment \cite{Kuzmenko2009,Tang2010}. The calculated phonon linewidth without field reproduces the features found in the previous DFT \cite{Park2008} and analytical calculations \cite{Ando2007} and agree reasonably well with the experimental data \cite{Yan2008}. The evolution of $\gamma$ for the $E_g$ mode as a function of $E$ also agrees quantitatively with that by Ando and Koshino \cite{Ando2009}.

For a given field $E$, the linewidths of the two modes change dramatically with the doping level $E_F$. This can be understood from the selective coupling between the phonon modes and electronic bands. In the low doping regime with $|E_F| < \hbar \omega_0/2 \sim 0.1$ eV, the low-branch mode can be in a resonant coupling with the electron-hole pair from the top valence band (v1) and the bottom conduction band (c1), as shown in Fig.~4. As a result, the linewidth of the low-branch mode is a constant within this doping range, as depicted in Fig.~\ref{fig:gamma}(a). Beyond this range, the linewidth of $E_g$ mode becomes nearly zero, while $\gamma$ of the $E_u$ mode increases. This is because $E_u$ mode couples with the two valence bands (v1 and v2) and conduction bands (c1 and c2), respectively, which has been discussed in detail in our previous work \cite{me2012}. This feature is not captured by Ref. [\onlinecite{Ando2009}] because of their simplified band structure of BLG. In view of the slight mixing effect on the phonon mode, we conclude that it is mainly the band structure effect that plays a dominant role in the tunable interplay between electrons and phonons in BLG induced by EEF and $E_F$.

The corresponding frequency shifts $\Delta \omega$ are presented in Figs.~\ref{fig:gamma}(c) and~\ref{fig:gamma}(d), respectively. In particular, the low-branch $E_g$ mode exhibits a frequency increase (hardening) when the $|E_F| > \hbar\omega_0$/2. In contrast, the high-branch $E_u$ mode softens with the increase of $|E_F|$. As a result, two distinguishable peaks will appear in the Raman or IR spectra of gated BLG. Since the frequency of the $E_u$ mode (1592 cm$^{-1}$) is higher than the $E_g$ mode (1587 cm$^{-1}$), one expects a crossing of the two branches as $|E_F|$ increases. This is in agreement with the Raman data by Yan et al. \cite{Yan2009}. From Figs.~\ref{fig:gamma}(c) and~\ref{fig:gamma}(d), for a higher $E$, the $E_g$ mode can be tuned by $|E_F|$ and hardens even faster, implying an interesting interplay between EEF, electrons and phonons in BLG.

In summary, we found that the electric field only slightly mixes the two in-plane modes, while inducing a relatively large change on the out-of-plane modes. Because of the broken symmetry, the two in-plane modes become Raman and IR active, and are detectable by both Raman and IR spectroscopy. The delicate electron-phonon coupling can be tuned through doping and an external gate field. It is mainly the band structure of BLG, tunable by external field and charge carrier densities, that dominates the dependence of the phonon linewidth and frequency shift as a function of $E_F$ and $E$. Our findings have revealed the external field effects on the phonon frequency renormalization observed in experiment, and may also be useful to study the field effects on other layered systems such as multilayer graphene.

J.A.Y. thanks Mei-Yin Chou and Kalman Varga for fruitful discussions. This work used the computer resources of Carver at NERSC and Kraken at the National Institute for Computational Sciences under an XSEDE startup allocation (Request No. DMR110111).


\begin{thebibliography}{99}
\bibitem{note} The phonon eigenvectors consist of three components ($x$, $y$, $z$) for each of the four atoms in the unit cell of BLG (as indicated in Figs. 1 and 2). For the doubly degenerate in-plane modes, we only consider the non-zero $x$ vibrations. All the $y$ and $z$ components are suppressed since they are zero. Similarly, we suppress the $x$ and $y$ components for the out-of-plane modes below.
\end{thebibliography}
\end{document}